\documentclass[12pt]{article}

\usepackage{graphicx}

\evensidemargin\oddsidemargin
\newdimen\picwidth

\def\ibar{\overline I_0}
\def\qsqmax{q^2_{\mathrm{max}}}
\def\gres{g_{\mathrm{res}}}
\def\mres{m_{\mathrm{res}}}

\def\dotp{\mathord{\cdot}}
\def\kev{\,\mathrm{ke\kern-0.1em V}}
\def\mev{\,\mathrm{Me\kern-0.1em V}}
\def\gev{\,\mathrm{Ge\kern-0.1em V}}
\def\tev{\,\mathrm{Te\kern-0.1em V}}
\def\re{\mathop{\mathrm{Re}}}

\def\n#1e#2n{{#1}\times 10^{#2}}

\newcommand{\err}[2]{%
\mathbin{\raisebox{0.05ex}[0pt][0pt]{\renewcommand{\arraystretch}{0}%
$\begin{array}{@{}c@{}}+\\-\end{array}$}}%
\raisebox{0.1ex}[0pt][0pt]{\renewcommand{\arraystretch}{0.35}\footnotesize%
$\begin{array}{@{}r@{}}#1\\#2\end{array}$}%
}

\makeatletter
\long\def\@makecaption#1#2{%
  \vskip\abovecaptionskip
  \sbox\@tempboxa{\small{\bfseries #1} \  #2}%
  \ifdim \wd\@tempboxa >\hsize
    \small{\bfseries #1} \  #2\par
  \else
    \global \@minipagefalse
    \hb@xt@\hsize{\hfil\box\@tempboxa\hfil}%
  \fi
  \vskip\belowcaptionskip}
\makeatother

\begin{document}

\begin{flushright}
hep-ph/0007263\\
SHEP 00 10
\end{flushright}

\begin{center}
{\Large\bfseries
Form Factors for Semileptonic $B\to\pi$ and $D\to\pi$ Decays from
the Omn\`es Representation}\\[2ex]
JM~Flynn$^{\mathrm{a}}$ and J~Nieves$^{\mathrm{b}}$\\[2ex]
${}^{\mathrm{a}}$Department of Physics \& Astronomy, University of
Southampton, Southampton SO17 1BJ, UK\\
${}^{\mathrm{b}}$Departamento de F\'\i sica Moderna, Universidad de
Granada, E--18071 Granada, Spain
\end{center}

\begin{abstract}\noindent
We use the Omn\`es representation to obtain the $q^2$ dependence of
the form factors $f^{+,0}(q^2)$ for semileptonic $H\to\pi$ decays from
elastic $\pi H \to\pi H$ scattering amplitudes, where $H$ denotes a
$B$ or $D$ meson. The amplitudes used satisfy elastic unitarity and
are obtained from two-particle irreducible amplitudes calculated in
tree-level heavy meson chiral perturbation theory (HMChPT). The
$q^2$-dependences for the form factors agree with lattice QCD results
when the HMChPT coupling constant, $g$, takes values smaller than
$0.32$, and confirm the milder dependence of $f^0$ on $q^2$ found in
sumrule calculations.
\end{abstract}

\section{Introduction} 

In this letter we present a description of the form factors $f^+$ and
$f^0$ describing semileptonic $H\to\pi$ decays, where $H$ denotes a
$D$ or $B$ meson. For the $B$ meson this exclusive semileptonic decay
can be used to determine the magnitude of the CKM matrix element
$V_{ub}$, currently the least well-known entry in the CKM
matrix. Ultimately, experimental measurements of $f_B^+(q^2)$ for
given momentum-transfer $q$ will be compared directly to theoretical
determinations at the same $q^2$ values to determine $|V_{ub}|$. In
the interim, it may be helpful to consider the decay rate integrated
partially or completely over $q^2$, but this requires knowledge of the
$q^2$ dependence of the form factors. Lattice calculations and sumrule
calculations apply in (different) restricted ranges of $q^2$ while
dispersion relations may be used to bound the form factors over the
whole $q^2$ range~\cite{bgl,lpl-bounds}, or as a basis for
models~\cite{buka}. A variety of models exists for the whole range of
$q^2$. One can ensure that general kinematic relations and the demands
of heavy quark symmetry (HQS) are satisfied, but an ansatz, such as
pole, dipole or other forms, is still required~\cite{burford,bkform}.

Here we use the Omn\`es representation to obtain the full $q^2$
dependence of these form factors from the elastic $\pi H \to \pi H$
scattering amplitudes.  For our application we have an isospin-$1/2$
channel, with angular momentum $J=1$ or $0$ for $f^+$ and $f^0$
respectively.  We rely on the following description of the (inverse)
amplitude for elastic $\pi H \to \pi H$ scattering in the isospin~$I$,
angular momentum~$J$, channel, with centre-of-mass squared-energy $s$
and masses $m$ and $M$ respectively~\cite{jnera},
\begin{equation}
\label{eq:tinv}
T_{IJ}^{-1}(s) = -\ibar(s) -C_{IJ} + 1/V_{IJ}(s),
\end{equation}
where $V_{IJ}$ is the two-particle irreducible scattering amplitude
and $C_{IJ}$ is a constant. $C_{IJ}$ and $V_{IJ}$ are real in the
scattering region.  This description implements elastic unitarity
automatically. Equation~(\ref{eq:tinv}) is justified by a dispersion
relation for $T^{-1}$, where the contributions of the left hand cut
and the poles (if any) are contained in $-C_{IJ} + 1/V_{IJ}$. $\ibar$
gives the exact contribution from the right hand cut, after any
necessary subtractions\footnote{$\ibar$ is calculated from a one-loop
`bubble' diagram. In the notation of reference~\cite{I&Z}, $\ibar(s) =
T_G((m+M)^2)-T_G(s)$, where $M$ and $m$ are the masses of the two
propagating particles}.  The description of equation~(\ref{eq:tinv})
may also be justified by an approach using the Bethe--Salpeter
equation.

Once $T_{IJ}$ is known, we can compute the corresponding phase shift
$\delta_{IJ}$. In turn, $\delta_{IJ}$ can be used in an Omn\`es
representation~\cite{omnes} giving $f_{IJ}(q^2)/ f_{IJ}(0)$ in terms
of an integral involving the phase shift, assuming that at threshold
the phase shift should be $n \pi$, where $n$ is the number of bound
states in the particular channel considered, and $\delta_{IJ} (\infty)
= k \pi$, where $k$ is the number of zeros of the scattering amplitude
on the physical sheet (this is Levinson's theorem~\cite{levinson}).

We determine $V_{IJ}$ from tree level heavy meson chiral perturbation
theory (HMChPT) \cite{HMChPT}, which implements HQS and is a double
expansion in powers of $1/M$ and momenta, where $M$ is the heavy meson
mass. The parameter $C_{IJ}$ in equation~(\ref{eq:tinv}) partially
accounts for higher order contributions in the expansion~\cite{jnera}.

We find consistency of our description with lattice results for the
$D\to\pi$ \cite{ukqcd:dtok,ape2000} and $B \to \pi$
\cite{ape2000,ukqcd:btopi} form factors if we set the HMChPT coupling,
$g$, to values smaller than $0.32$. This upper bound is in reasonable
agreement with other determinations, but $g$ is not very well known
\cite{hmchptreview,ukqcd:gbstarbpi}.

Our model and the Omn\`es representation are not guaranteed at high
energies where inelasticities become important. However, our
hypothesis is that only the low-lying states and energies should
influence the form factors we consider.

A dispersive approach to the $f^+$ form factor was taken by Burdman
and Kambor~\cite{buka}, who also used HMChPT to calculate the phase
shift in $\pi H\to \pi H$ scattering. Here by working with the
\emph{inverse} amplitude we can ensure that Watson's theorem and
elastic unitarity are satisfied exactly. Moreover, we compute $f^+$
and $f^0$ together to examine whether different behaviours in $q^2$
are found, consistent with lattice QCD results and allowing extra
information from $f^0$ to be used to constrain $f^+$.

\section{Scattering Amplitudes and Form Factors}

We compute $V_{1/2}$, the two-particle irreducible amplitude for $\pi
H$ scattering in the isospin $1/2$ channel, $\pi(p_1) H(Mv) \to
\pi(p_2) H(Mv+q_2)$. Here, $v$ is the four-velocity of the initial
heavy meson of mass $M$. The pion mass is $m$. We use the direct tree
level interaction from the lowest order HMChPT lagrangian, together
with tree diagrams for $H^*$ exchange which involve the leading
interaction term with coupling $g$ \cite{HMChPT,hmchptreview}. The
result is,
\begin{equation}
V_{1/2} = -{M\over f^2}
 \left\{ 3 v\dotp p_1 + v\dotp p_2 +
         g^2 (p_1\dotp p_2 - v\dotp p_1 v\dotp p_2)
         \left( {3\over v\dotp p_1 - \Delta} +
                {1\over v\dotp p_2 + \Delta}
         \right)
 \right\}.
\end{equation}
Here, $f = 130.7\mev$ is the pion decay constant and $\Delta =
(M_{*}^2-M^2)/2M \approx M_* - M$, where $M_*$ is the heavy vector
meson mass. We subsequently project $V_{1/2}$ onto the angular
momentum $0$ and $1$ channels.

The full scattering amplitude at centre of mass energy-squared $s$, in
the isospin $I$ and angular momentum $J$ channel, is obtained in our
approach from equation~(\ref{eq:tinv}).  The phase shift $\delta_{IJ}$
is then obtained from,
\begin{equation}
T_{IJ}(s) = {8\pi i s\over \lambda^{1/2}(s,M^2,m^2)} \,
         (e^{2i\delta_{IJ}(s)}-1),
\end{equation}
where $\lambda(x,y,z) = x^2 + y^2 + z^2 - 2(xy + yz + zx)$ is the
usual kinematic function.

Once the phase shift is known, we use the Omn\`es representation to
obtain the $q^2$ dependence of the form factors as follows:
\begin{equation}
\label{eq:omnes}
{f(q^2)\over f(0)} =
 \exp\left[
  {q^2\over\pi}
  \int_{(m+M)^2}^\infty {\delta_{IJ}(s) \, ds\over s (s-q^2)}
 \right].
\end{equation}
In this work, we always have $I=1/2$. The form factor $f^+$ is
obtained when $J=1$ and depends on $J^P = 1^-$ resonances, while $f^0$
is obtained when $J=0$ and depends on $J^P = 0^+$ resonances. We
perform the integral numerically, taking the upper limit as $100$
times the lower limit\footnote{$f^+_{D\pi}(\qsqmax)$, where $\qsqmax =
(m_D-m_\pi)^2$, varies by less than $1\%$ as the upper limit of
integration varies from 50 to 200 times the lower limit, and the
variation is smaller at lower $q^2$.}. The form factors are equal at
$q^2=0$: $f^+(0)=f^0(0)$.

\begin{itemize}
\item 
For $J^P = 1^-$ we take $C=0$ for the $D$ decay because the $D^*$
resonance is so close to threshold that we expect it to saturate all
the counterterms in HMChPT (compare to vector meson dominance in
$\pi\pi$ scattering in ordinary chiral perturbation
theory). Calculating $C$ in this case reveals the value $C=8\times
10^{-6}.$ We still have the freedom to vary the lowest order coupling
constant $g$ in HMChPT. For the $B$ meson decay, we set $C=-0.0014$ to
keep the $B^*$ pole at its correct mass.
\item
For $J^P = 0^+$ we ignore $D^*$ and $B^*$ $s$-channel exchanges, which
have the wrong quantum numbers to contribute in this case. These
exchanges only contribute because of the heavy meson mass
expansion implicit in HMChPT. Instead we
keep $C$ non-zero, setting $C = -0.0051$ for the $D$-physics case to
get a resonance at about $2350\mev$, and $C = -0.0016$ for
$B$-physics to get a resonance at about $5660\mev$~\cite{ciulliHF8}.
\end{itemize}

The values of $C$ are determined by demanding that $T^{-1}$ ($\re
T^{-1}$) vanishes at the position of a pole (resonance). For the $J=1$
channels, $V^{-1}$ vanishes by construction at the positions of the
$D^*$ or $B^*$, and so, from equation~(\ref{eq:tinv}), $C$ is
independent of $g$. In the $J=0$ channels, $g$-dependence enters in
$V^{-1}$, but only through the $t$-channel tree graphs, and is very
weak. $C$ varies by less than $0.5\%$ for $0<g<0.45$ in the $D$-meson
case and the dependence is even weaker for the $B$-meson case.

We noted that in using the Omn\`es representation~\cite{omnes} of
equation~(\ref{eq:omnes}), the phase shift at threshold should be
$n\pi$, where $n$ is the number of bound states in the channel under
consideration. Thus $n=0$ in all channels used
here except for $J^P = 1^-$ in the $B$ case where $n=1$ to account for
the $B^*$. In fact, our model also gives a bound state in the $0^+$
channel in the $B$ case, which we ignore. One could try to improve the
model to avoid this unphysical bound state by replacing $C$ with a function of
$q^2$ (the function should have no right hand cut).

\section{Semileptonic Decays}

The process $D^*\to D\pi$ is kinematically allowed, so the $D^*$ is a
resonance in $D\pi$ scattering. In HMChPT the decay rates of
$D^{*\,+}$ to $D^0\pi^+$ and $D^+\pi^0$ are given to lowest order by,
\begin{equation}
\Gamma(D^{*\,+}\to D^0 \pi^+) = {g^2 p^3\over 6 \pi f^2},
\qquad
\Gamma(D^{*\,+}\to D^+ \pi^0) = {g^2 p^3\over 12 \pi f^2}.
\end{equation}
The sum of these rates can also be obtained from the slope of the
phase shift at the resonance mass. We find that these two methods
agree for a range of $g$ values.

\begin{figure}
\hbox to\hsize{\hss
\small
\picwidth=\hsize
\setlength{\unitlength}{\picwidth}\divide\unitlength by5400%
\begin{picture}(5400,1620)(0,0)%
\put(0,0){\includegraphics[width=\picwidth]{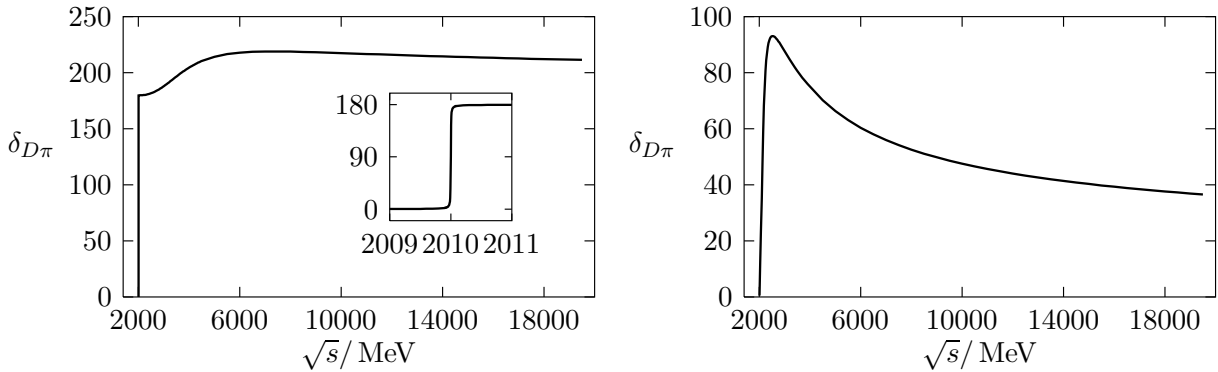}}
\put(4225,50){\makebox(0,0){$\sqrt{s}/\mev$}}%
\put(2800,910){\makebox(0,0)[b]{\shortstack{$\delta_{D\pi}$}}}%
\put(5030,200){\makebox(0,0){$18000$}}%
\put(4589,200){\makebox(0,0){$14000$}}%
\put(4148,200){\makebox(0,0){$10000$}}%
\put(3707,200){\makebox(0,0){$6000$}}%
\put(3266,200){\makebox(0,0){$2000$}}%
\put(3150,1520){\makebox(0,0)[r]{$100$}}%
\put(3150,1276){\makebox(0,0)[r]{$80$}}%
\put(3150,1032){\makebox(0,0)[r]{$60$}}%
\put(3150,788){\makebox(0,0)[r]{$40$}}%
\put(3150,544){\makebox(0,0)[r]{$20$}}%
\put(3150,300){\makebox(0,0)[r]{$0$}}%
\put(2190,532){\makebox(0,0){$2011$}}%
\put(1925,532){\makebox(0,0){$2010$}}%
\put(1659,532){\makebox(0,0){$2009$}}%
\put(1609,1137){\makebox(0,0)[r]{$180$}}%
\put(1609,910){\makebox(0,0)[r]{$90$}}%
\put(1609,682){\makebox(0,0)[r]{$0$}}%
\put(1525,50){\makebox(0,0){$\sqrt{s}/\mev$}}%
\put(100,910){\makebox(0,0)[b]{\shortstack{$\delta_{D\pi}$}}}%
\put(2330,200){\makebox(0,0){$18000$}}%
\put(1889,200){\makebox(0,0){$14000$}}%
\put(1448,200){\makebox(0,0){$10000$}}%
\put(1007,200){\makebox(0,0){$6000$}}%
\put(566,200){\makebox(0,0){$2000$}}%
\put(450,1520){\makebox(0,0)[r]{$250$}}%
\put(450,1276){\makebox(0,0)[r]{$200$}}%
\put(450,1032){\makebox(0,0)[r]{$150$}}%
\put(450,788){\makebox(0,0)[r]{$100$}}%
\put(450,544){\makebox(0,0)[r]{$50$}}%
\put(450,300){\makebox(0,0)[r]{$0$}}%
\end{picture}%
\hss}
\caption[]{Phase $\delta_{D\pi}$ for the $IJ=1/2,1$ (left) and $1/2,0$
(right) channels in $D\pi$ scattering. The inset on the left shows the
resonance at $\sqrt{s}=m_{D^*}=2010\mev$. Phases are calculated with
$g=0.21$.}
\label{fig:phaseD}
\end{figure}

The $D^*$ exchange is included in our tree level amplitude, and we
expect it to saturate the counterterms in HMChPT, so in calculating
$T_{1/2,1}$ we set $C=0$ as noted above. Figure~\ref{fig:phaseD}
(left) shows the phase shift obtained for $J=1$. With input masses,
$m_D = 1864.5\mev$ and $m_{D^*} = 2010\mev$, the $D^*$ resonance shows
up as the jump of $\pi$ in the phase at the $D^*$ mass.  In the $J=0$
channel, we tune $C$ to produce a resonance at the expected mass of
the $D_0^*$ at $2350\mev$ \cite{ciulliHF8}. The $J=0$ phase shift is
shown on the right in figure~\ref{fig:phaseD}.

\begin{figure}
\hbox to\hsize{\hss
\small
\picwidth=\hsize
\setlength{\unitlength}{\picwidth}\divide\unitlength by5400%
\begin{picture}(5400,1620)(0,0)%
\put(0,0){\includegraphics[width=\picwidth]{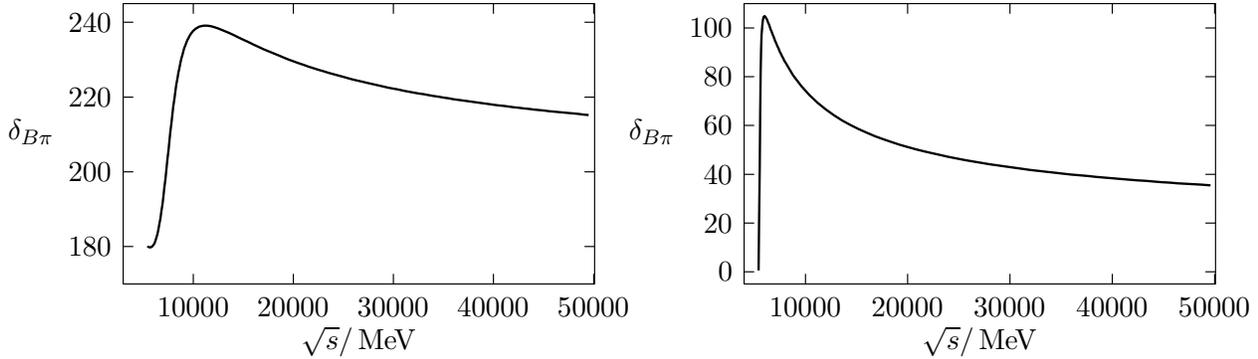}}
\put(4225,50){\makebox(0,0){$\sqrt{s}/\mev$}}%
\put(2800,910){\makebox(0,0)[b]{\shortstack{$\delta_{B\pi}$}}}%
\put(5246,200){\makebox(0,0){$50000$}}%
\put(4801,200){\makebox(0,0){$40000$}}%
\put(4356,200){\makebox(0,0){$30000$}}%
\put(3911,200){\makebox(0,0){$20000$}}%
\put(3467,200){\makebox(0,0){$10000$}}%
\put(3150,1414){\makebox(0,0)[r]{$100$}}%
\put(3150,1202){\makebox(0,0)[r]{$80$}}%
\put(3150,990){\makebox(0,0)[r]{$60$}}%
\put(3150,777){\makebox(0,0)[r]{$40$}}%
\put(3150,565){\makebox(0,0)[r]{$20$}}%
\put(3150,353){\makebox(0,0)[r]{$0$}}%
\put(1525,50){\makebox(0,0){$\sqrt{s}/\mev$}}%
\put(100,910){\makebox(0,0)[b]{\shortstack{$\delta_{B\pi}$}}}%
\put(2546,200){\makebox(0,0){$50000$}}%
\put(2110,200){\makebox(0,0){$40000$}}%
\put(1675,200){\makebox(0,0){$30000$}}%
\put(1240,200){\makebox(0,0){$20000$}}%
\put(805,200){\makebox(0,0){$10000$}}%
\put(450,1439){\makebox(0,0)[r]{$240$}}%
\put(450,1113){\makebox(0,0)[r]{$220$}}%
\put(450,788){\makebox(0,0)[r]{$200$}}%
\put(450,463){\makebox(0,0)[r]{$180$}}%
\end{picture}%
\hss}
\caption[]{Phase $\delta_{B\pi}$ for the $IJ=1/2,1$ (left) and
$IJ=1/2,0$ (right) channels in $B\pi$ scattering. Phases are
calculated with $g=0.21$.}
\label{fig:phaseB}
\end{figure}

In the $B$ case, the decay process $B^*\to B\pi$ is not kinematically
allowed and the $B^*$ meson is a pole, sitting between the maximum
physical $q^2$ value for the form factor, $\qsqmax = (m_B - m_\pi)^2$,
and the start of the physical cut at $q^2 = (m_B + m_\pi)^2$. Again we
use the physical pseudoscalar and vector meson masses as inputs, $m_B
= 5278.9\mev$, $m_{B^*} = 5324.8\mev$. The phase shift for the $J=1$
case is shown on the left in figure~\ref{fig:phaseB}. The appearance
of the $B^*$ as a bound state between $\qsqmax$ and $q^2 = (m_B +
m_\pi)^2$ is signalled by the vanishing of $T^{-1}_{1/2,1}(s)$ at
$\sqrt s=m_{B^*}$. The $J=0$ phase shift appears on the right of
figure~\ref{fig:phaseB}.

\begin{figure}
\hbox to\hsize{\hss
\small
\picwidth=\hsize
\setlength{\unitlength}{\picwidth}\divide\unitlength by5400%
\begin{picture}(5400,1620)(0,0)%
\put(0,0){\includegraphics[width=\picwidth]{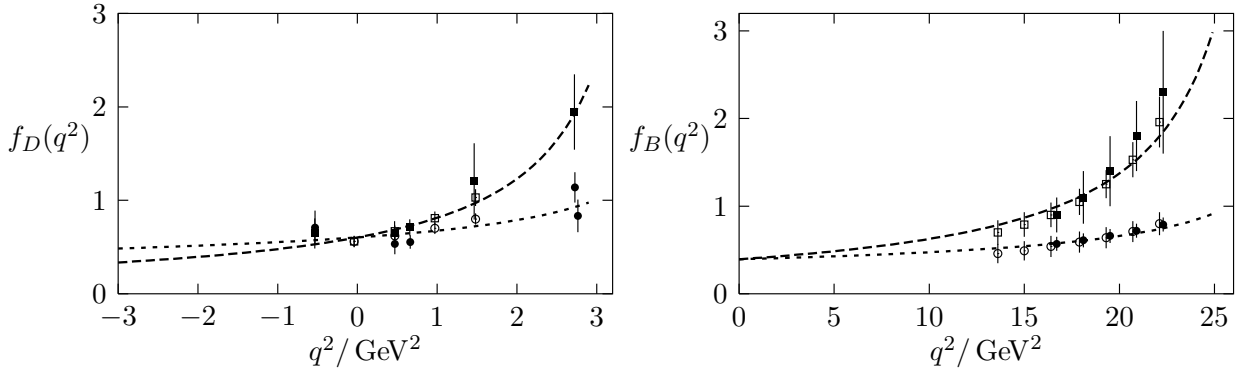}}
\put(4175,50){\makebox(0,0){$q^2/\gev^2$}}%
\put(2800,910){\makebox(0,0)[b]{\shortstack{$f_B(q^2)$}}}%
\put(5167,200){\makebox(0,0){$25$}}%
\put(4754,200){\makebox(0,0){$20$}}%
\put(4340,200){\makebox(0,0){$15$}}%
\put(3927,200){\makebox(0,0){$10$}}%
\put(3513,200){\makebox(0,0){$5$}}%
\put(3100,200){\makebox(0,0){$0$}}%
\put(3050,1444){\makebox(0,0)[r]{$3$}}%
\put(3050,1063){\makebox(0,0)[r]{$2$}}%
\put(3050,681){\makebox(0,0)[r]{$1$}}%
\put(3050,300){\makebox(0,0)[r]{$0$}}%
\put(1475,50){\makebox(0,0){$q^2/\gev^2$}}%
\put(100,910){\makebox(0,0)[b]{\shortstack{$f_D(q^2)$}}}%
\put(2481,200){\makebox(0,0){$3$}}%
\put(2134,200){\makebox(0,0){$2$}}%
\put(1787,200){\makebox(0,0){$1$}}%
\put(1440,200){\makebox(0,0){$0$}}%
\put(1094,200){\makebox(0,0){$-1$}}%
\put(747,200){\makebox(0,0){$-2$}}%
\put(400,200){\makebox(0,0){$-3$}}%
\put(350,1520){\makebox(0,0)[r]{$3$}}%
\put(350,1113){\makebox(0,0)[r]{$2$}}%
\put(350,707){\makebox(0,0)[r]{$1$}}%
\put(350,300){\makebox(0,0)[r]{$0$}}%
\end{picture}%
\hss}
\caption[]{Form factors in $D\to\pi$ (left) and $B\to\pi$ (right)
semileptonic decays. The squares (circles) denote $f^+$ ($f^0$)
from lattice calculations, while the long-dashed (short-dashed) lines
denote the fitted curves for $f^+$ ($f^0$). Solid symbols are results
from UKQCD~\cite{ukqcd:dtok,ukqcd:btopi}, open symbols are results
from APE~\cite{ape2000}.}
\label{fig:form-factors}
\end{figure}

\begin{figure}
\hbox to\hsize{\hss
\small
\picwidth=\hsize
\setlength{\unitlength}{\picwidth}\divide\unitlength by5400%
\begin{picture}(5400,1620)(0,0)%
\put(0,0){\includegraphics[width=\picwidth]{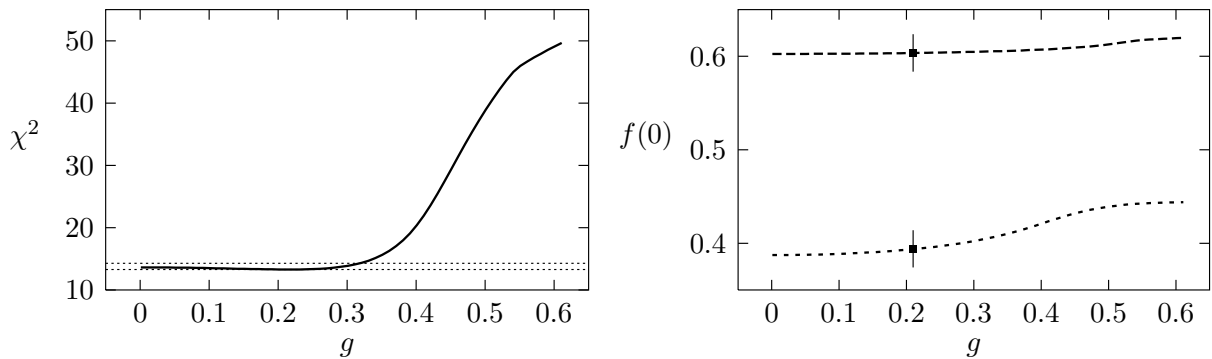}}
\put(4225,50){\makebox(0,0){$g$}}%
\put(2800,910){\makebox(0,0)[b]{\shortstack{$f(0)$}}}%
\put(5104,200){\makebox(0,0){$0.6$}}%
\put(4811,200){\makebox(0,0){$0.5$}}%
\put(4518,200){\makebox(0,0){$0.4$}}%
\put(4225,200){\makebox(0,0){$0.3$}}%
\put(3932,200){\makebox(0,0){$0.2$}}%
\put(3639,200){\makebox(0,0){$0.1$}}%
\put(3346,200){\makebox(0,0){$0$}}%
\put(3150,1317){\makebox(0,0)[r]{$0.6$}}%
\put(3150,910){\makebox(0,0)[r]{$0.5$}}%
\put(3150,503){\makebox(0,0)[r]{$0.4$}}%
\put(1500,50){\makebox(0,0){$g$}}%
\put(100,910){\makebox(0,0)[b]{\shortstack{$\chi^2$}}}%
\put(2400,200){\makebox(0,0){$0.6$}}%
\put(2100,200){\makebox(0,0){$0.5$}}%
\put(1800,200){\makebox(0,0){$0.4$}}%
\put(1500,200){\makebox(0,0){$0.3$}}%
\put(1200,200){\makebox(0,0){$0.2$}}%
\put(900,200){\makebox(0,0){$0.1$}}%
\put(600,200){\makebox(0,0){$0$}}%
\put(400,1384){\makebox(0,0)[r]{$50$}}%
\put(400,1113){\makebox(0,0)[r]{$40$}}%
\put(400,842){\makebox(0,0)[r]{$30$}}%
\put(400,571){\makebox(0,0)[r]{$20$}}%
\put(400,300){\makebox(0,0)[r]{$10$}}%
\end{picture}%
\hss}
\caption[]{Left: chi-squared for the fit described in the text as a
function of the HMChPT coupling $g$, for $g\geq0.001$. Right: values
of $f^{+,0}_B(0)$ (lower curve) and $f^{+,0}_D(0)$ (upper curve) as
functions of $g$. The points with errors on the right are the best fit
values of equation~(\ref{eq:res}) at $g=0.21$.}
\label{fig:depeng}
\end{figure}

From the phase shifts we find the form factors $f^+$ and $f^0$. We
perform a simultaneous three-parameter fit to the UKQCD and APE
lattice results~\cite{ukqcd:dtok,ape2000,ukqcd:btopi} for the form
factors $f^+(q^2)$ and $f^0(q^2)$ which determine the $B$ and $D$
semileptonic decays. The free parameters are the HMChPT coupling
constant $g$ and the form factors at $q^2=0$: $f_B(0)$ for $B \to \pi$
decays and $f_D(0)$ for $D \to \pi$ decays\footnote{To use the results
in~\cite{ukqcd:dtok}, we take $Z_V^{\rm eff}=0.88$ for the vector
renormalisation constant connecting lattice and continuum
results.}. The best fit parameters with 39 degrees of freedom are
\begin{equation}
g = 0.21 \err{0.11}{0.21}, \quad f_B(0) = 0.39 \pm 0.02, \quad f_D(0) =
0.60 \pm 0.02 \quad\mathrm{with}\quad \chi^2/\mathrm{dof} = 0.34.
\label{eq:res}
\end{equation}
Results can be seen in figure~\ref{fig:form-factors}. Errors in the
fitted parameters are statistical and have been obtained by increasing
the value of the total $\chi^2$ by one unit. A word of caution must be
stated about the results for the HMChPT coupling constant $g$. Scalar
channels are almost insensitive to this parameter. For the vector
channels, in the case of $D$ meson decay, the resonance is so close to
threshold that it completely dominates the process, independent of the
value of $g$, as long as the resonant contribution is more important
than the background. This turns out to be true as long as $g$ is
greater than $0.001$, thus the smallest value $g$ can take is $0.001$
and not zero as can be inferred from equation~(\ref{eq:res}). To
clarify the dependence of our results on $g$, we show in
figure~\ref{fig:depeng} both $\chi^2$ and $f_B(0)$, $f_D(0)$ versus
$g$, for $g\geq0.001$. In the first figure the line at $\chi^2 =
13.28$ shows the minimum value of $\chi^2$, while the line at $\chi^2
= 14.28$ determines the upper error. We also show best fit values,
with fixed $g$, of $f_B(0)$ and $f_D(0)$ versus $g$. The points with
errors correspond to the results quoted in equation~(\ref{eq:res}).

We note that $f^+_D$ is well-approximated by a simple pole form with
the $D^*$ giving the pole mass, while $f^0_D$ is noticeably `flatter'
in $q^2$. This is consistent with lattice results. For the $B\to\pi$
case, $f^+_B$ is well-approximated by a pole form with the pole mass
of order the $B^*$ meson mass. The $f^0_B$ form factor has much less
$q^2$ dependence, consistent with the behaviour found in lattice
calculations.

We have also determined the coupling $g$ and form factors at $q^2=0$
separately for $D$ and $B$ decays using independent fits to the UKQCD
and APE lattice data for $D$ and $B$. The best fit values turn out to
be the same as in equation~(\ref{eq:res}), although $g_D$ can be as
large as $0.46$ while still giving an acceptable chi-squared.  To
compare with light cone sumrule (LCSR) results, we take the LCSR
values $f_{D^*}g_{D^*D\pi} = 2.7 \pm 0.8 \gev$ and $f_{B^*}g_{B^*B\pi}
= 4.4 \pm 1.3 \gev$~\cite{krwwy}, and combine with lattice
calculations of the vector meson decay constants from Becirevic et
al~\cite{becir} and UKQCD~\cite{ukqcd:decay-consts-2000}, to yield
\begin{equation}
\label{eq:gDgB-lcsr}
g_D = \cases{0.35\err{0.12}{0.11}& $f_{D^*}$ from \cite{becir}\cr
             0.39\pm0.12& $f_{D^*}$ from
                \cite{ukqcd:decay-consts-2000}\cr}
\qquad
g_B = \cases{0.23\pm0.08& $f_{B^*}$ from \cite{becir}\cr
             0.28\err{0.10}{0.09}& $f_{B^*}$ from
                \cite{ukqcd:decay-consts-2000}\cr}
\end{equation}
The values are quite compatible in the $B$ case, less so for $D$
decays, although, as noted above, our fit for $g_D$ allowed a large
variation above the best-fit value. The value of $f_D(0)$ found here
agrees well with the LCSR result $f^+_D(0)=0.65\pm0.11$~\cite{krwwy},
while $f_B(0)$ in equation~(\ref{eq:res}) is higher than the LCSR
value $f^+_B(0) = 0.28 \pm 0.05$~\cite{krwwy}. In the $D$ case, the
$D^*$ resonance is only a few $\mev$ above threshold and the range of
$q^2$ for the semileptonic decay is not large, so one expects a simple
pole form for $f^+$ to work well. For $B$ physics, the effects
of higher resonances and continuum states are evidently more
important: such effects are incorporated in LCSR calculations but are
not present in the very simple model used here. We address this issue
in section~\ref{sec:extrares} below.


Heavy quark symmetry (HQS) is an input in HMChPT. The HQS scaling
relations for the $B$ decay form factors at $\qsqmax$ are preserved
because $f^+(\qsqmax)/f^0(\qsqmax)$ is proportional to
\begin{equation}
\exp\left({\qsqmax\over\pi}\, \int_{(M+m)^2}^\infty
 {\delta^+ - \delta^0 \over s(s-\qsqmax)} \, ds \right).
\end{equation}
The above result relies on the equality of the form factors at
$q^2=0$, $f^+(0)=f^0(0)$.  If $\delta^+ - \delta^0 = \pi$, which we
see is satisfied by our phase shifts at large $\sqrt{s}$, then the
ratio $f^+(\qsqmax)/f^0(\qsqmax) = 1/(1-\qsqmax/(M+m)^2)$ as demanded
by HQS, where $M$ and $m$ are the masses of the heavy meson and the
pion respectively.

We have applied the same approach to describe semileptonic $D \to K$
decays. Here, it gives form factors flatter than lattice
results~\cite{ukqcd:dtok} and the experimental evidence~\cite{CLEO}.
However, corrections of both types $m_K/m_D$ and $m_K^2/(4\pi
f_\pi)^2$ to the tree level HMChPT results used here are expected to
be sizeable in this case.

\section{Extra Resonances}
\label{sec:extrares}

\begin{figure}
\hbox to\hsize{\hss
\small
\picwidth=\hsize
\setlength{\unitlength}{\picwidth}\divide\unitlength by5400%
\begin{picture}(5400,1620)(0,0)%
\put(0,0){\includegraphics[width=\picwidth]{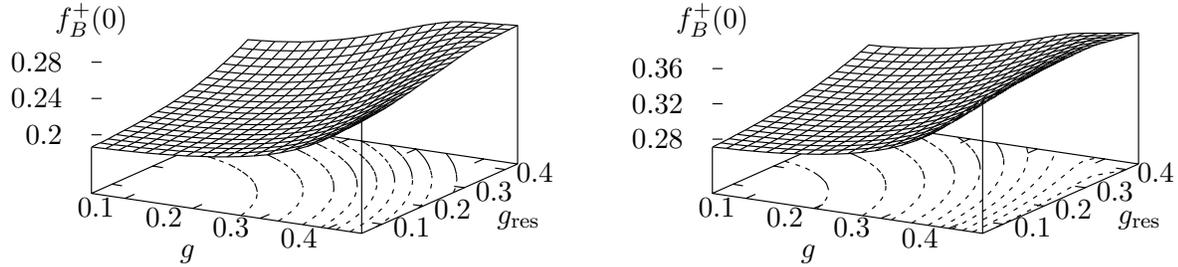}}
\put(3123,1248){\makebox(0,0){$f_B^+(0)$}}%
\put(2997,1037){\makebox(0,0)[r]{$0.36$}}%
\put(2997,884){\makebox(0,0)[r]{$0.32$}}%
\put(2997,730){\makebox(0,0)[r]{$0.28$}}%
\put(4980,378){\makebox(0,0){$\gres$}}%
\put(4971,583){\makebox(0,0)[l]{$0.4$}}%
\put(4793,504){\makebox(0,0)[l]{$0.3$}}%
\put(4614,424){\makebox(0,0)[l]{$0.2$}}%
\put(4436,345){\makebox(0,0)[l]{$0.1$}}%
\put(3541,232){\makebox(0,0){$g$}}%
\put(4105,301){\makebox(0,0)[r]{$0.4$}}%
\put(3811,345){\makebox(0,0)[r]{$0.3$}}%
\put(3518,388){\makebox(0,0)[r]{$0.2$}}%
\put(3224,432){\makebox(0,0)[r]{$0.1$}}%
\put(423,1248){\makebox(0,0){$f_B^+(0)$}}%
\put(297,1065){\makebox(0,0)[r]{$0.28$}}%
\put(297,907){\makebox(0,0)[r]{$0.24$}}%
\put(297,749){\makebox(0,0)[r]{$0.2$}}%
\put(2280,378){\makebox(0,0){$\gres$}}%
\put(2271,583){\makebox(0,0)[l]{$0.4$}}%
\put(2093,504){\makebox(0,0)[l]{$0.3$}}%
\put(1914,424){\makebox(0,0)[l]{$0.2$}}%
\put(1736,345){\makebox(0,0)[l]{$0.1$}}%
\put(841,232){\makebox(0,0){$g$}}%
\put(1405,301){\makebox(0,0)[r]{$0.4$}}%
\put(1111,345){\makebox(0,0)[r]{$0.3$}}%
\put(818,388){\makebox(0,0)[r]{$0.2$}}%
\put(524,432){\makebox(0,0)[r]{$0.1$}}%
\end{picture}%
\hss}
\caption[]{Variation of $f_B^+(0)$ with $g$ and $\gres$ where the
$B^*$ coupling is $g$ and a second resonance is added in the $J=1$
channel with coupling $\gres$. On the left the resonance mass is
$\mres=6100\mev$ and contours are plotted from $f^+_B(0) = 0.20$ to
$0.28$ in increments of $0.01$; on the right $\mres=8100\mev$ and
contours are plotted from $f^+_B(0) = 0.28$ to $0.37$ in increments of
$0.01$.}
\label{fig:surface}
\end{figure}

We noted above that our result for $f_B(0)$ in equation~(\ref{eq:res})
is higher than the LCSR value of around $0.28$, while our fit for
$f^+_B(q^2)$ is well-approximated by a simple pole form with pole mass
of order $m_{B^*}$. This suggests that deviations from $B^*$ pole
dominance can become significant at low $q^2$. This phenomenon was
also noted by Burdman and Kambor~\cite{buka} who implemented a
constrained dispersive model for $f^+_B$. Likewise, lattice results
have favoured dipole forms in fits to
$f^+_B$~\cite{burford,ape2000,ukqcd:btopi}.

To address this issue we have added a second resonance of mass $\mres$
by hand in the 2PI $J=1$ amplitude $V_{1/2,1}$, coupling it like the
$H^*$ but with its own coupling strength $\gres$. In the $D$-meson
case, we already had a good fit to the lattice results and a
consistent value for $f_D(0)$. If $\mres$ is large enough the extra
resonance does not disturb this picture. In the $B$ case, we can
easily make $f_B(0)$ smaller while still fitting lattice results at
large $q^2$. In figure~\ref{fig:surface} we show $f_B(0)$ as a
function of the couplings $g$ and $\gres$ for two choices of the extra
resonance mass, $\mres=6100\mev, 8100\mev$. The problem in this case
is that it is not possible to make a statistically acceptable fit to
$f_B^+$ and $f_B^0$ simultaneously. One could try to add an extra
resonance in the $J=0$ channel also, but while our choice of
$\mres=6100\mev$ for $J=1$ may be motivated by potential
models~\cite{ehq} or lattice results~\cite{nrqcdspec}, we do not know
whether or how to set the mass for additional $J=0$ resonances, having
already set the $C$ values to account for rather poorly known
resonances. This emphasises the importance of looking at $f^+$ and
$f^0$ together, even though $f^+$ is the experimentally accessible
form factor.

\section{Conclusion}

Our model is extremely simple, using only tree level HMChPT
information for the two particle irreducible amplitude $V_{IJ}$,
thereby incorporating only the first excited hadron state. Furthermore
we fix to a constant an allowed polynomial in $q^2$ muliplying the
Omn\`es exponential factor in equation~(\ref{eq:omnes}). Thus,
deviations from LCSR results for $f^+$ are not unexpected because
those calculations incorporate effects of higher resonances and
continuum states.  Taking our model beyond leading order is not
possible at present because of the proliferation of undetermined
parameters which would appear in the next order of HMChPT and the lack
of experimental data to fix them. This is a standard difficulty in
using effective theories at higher orders.

The simple model presented here gives an excellent description of
semileptonic $D$-decays. For $B$-decays it gives a good description of
the lattice data near $\qsqmax$ and is also compatible within two
standard deviations with LCSR predictions at $q^2=0$. Moreover, it
provides a framework compatible with heavy quark symmetry, naturally
accomodating pole-like behaviour for $f^+$ and, simultaneously,
non-constant behaviour for $f^0$. Previously, as pointed out
in~\cite{burford,lpl-bounds}, a difficulty for form factor models with
pole-type behaviour for $f^+$ was fixing a behaviour for $f^0$ which
satisfied both the relation $f^0(0) = f^+(0)$ and the requirements of
heavy quark symmetry. Pole-like behaviour of $f^+$ turns out again to
be feasible in our model, thanks to the fact that the $B^*$ is a bound
state rather than a $\pi B$ resonance.

Qualitatively, the results found here are encouraging. However, the
larger value found for $f^+_B(0)$ compared to that from LCSR
calculations would lead to appreciably smaller values for
$|V_{ub}|$. We caution the reader that this should not be taken to
indicate a large theoretical spread in the value of $|V_{ub}|$ from
exclusive semileptonic $B\to\pi$ decays: one should bear in mind the
simplicity of the model used. We indicated how a second resonance in
the $J=1$ channel can restore compatibility with both LCSR and lattice
results for $f_B^+$, although this shifts the problem to making
$f_B^0$ compatible with the lattice data in a combined fit and
emphasises the importance of using information from both form factors.

\subsection*{Acknowledgements}
JN acknowledges support under grant DGES PB98--1367 and by the Junta
de Andaluc\'{\i}a FQM0225, and thanks the SHEP group for their
hospitality during part of this work. JMF acknowledges PPARC for
support under grant PPA/G/O/1998/00525. We thank E~Ruiz-Arriola
for useful discussions and G~Burdman for communications.

\end{document}